\documentclass[aps,reprint]{revtex4-1}

\usepackage{graphicx}
\usepackage{dcolumn}
\usepackage{natbib,aas_macros}
\usepackage{color}
\usepackage{amssymb} 
\usepackage{amsthm}
\usepackage{mathrsfs}
\usepackage{amsmath}
\usepackage{bbm}
\usepackage{hyperref}
\usepackage{lineno}
\usepackage{tikz}
\usepackage[normalem]{ulem}
\usepackage{algorithm}
\usepackage{algorithmicx}
\usepackage{algpseudocode}
\usepackage{comment}
\usepackage{soul}

\usetikzlibrary{external}
\usetikzlibrary{positioning}
\newcommand{\mb}{\mathbf}
\newcommand{\bb}{\mathbb}

\newcommand{\R}{\bb R}
\newcommand{\E}{\bb E}

\renewcommand{\mathbf}{\boldsymbol}
\newcommand{\mr}{\mathrm}

\newcommand{\fnr}{\mathrm{FNR}}
\newcommand{\fpr}{\mathrm{FPR}}

\newtheorem{definition}{Definition}

\begin{document} 
\title{
Boosting the Efficiency of Parametric Detection with Hierarchical Neural Networks
}

\author{Jingkai Yan$^{1,4}$, Robert Colgan$^{2,4}$, John Wright$^{1,4}$, Zsuzsa M\'arka$^{4,5}$, Imre Bartos$^{6}$, and Szabolcs M\'arka$^{3,4}$\vspace{.1in}}

\address{$^1$Department of Electrical Engineering, Columbia University in the City of New York, 500 W. 120th St., New York, NY 10027, USA\\
$^2$Department of Computer Science, Columbia University in the City of New York, 500 W. 120th St., New York, NY 10027, USA\\
$^3$Department of Physics, Columbia University in the City of New York, 538 W. 120th St., New York, NY 10027, USA\\
$^4$Data Science Institute, Columbia University in the City of New York, 550 W. 120th St., New York, NY 10027, USA\\
$^5$Columbia Astrophysics Laboratory, Columbia University in the City of New York, 538 W. 120th St., New York, NY 10027, USA\\
$^6$Department of Physics, University of Florida, PO Box 118440, Gainesville, FL 32611-8440, USA
}

\begin{abstract}

Gravitational wave astronomy is a vibrant field that leverages both classic and modern data processing techniques for the understanding of the universe. Various approaches have been proposed for improving the efficiency of the detection scheme, with hierarchical matched filtering being an important strategy. Meanwhile, deep learning methods have recently demonstrated both consistency with matched filtering methods and remarkable statistical performance. In this work, we propose Hierarchical Detection Network (HDN), a novel approach to efficient detection that combines ideas from hierarchical matching and deep learning. The network is trained using a novel loss function, which encodes simultaneously the goals of statistical accuracy and efficiency. We discuss the source of complexity reduction of the proposed model, and describe a general recipe for initialization with each layer specializing in different regions. We demonstrate the performance of HDN with experiments using open LIGO data and synthetic injections, and observe with two-layer models a $79\%$ efficiency gain compared with matched filtering at an equal error rate of $0.2\%$. Furthermore, we show how training a three-layer HDN initialized using two-layer model can further boost both accuracy and efficiency, highlighting the power of multiple simple layers in efficient detection.

\end{abstract}
\maketitle

\section{Introduction}
\label{sec:intro}

The study of gravitational wave (GW) signals~\cite{1916SPAW.......688E,1918SPAW.......154E,PhysRevLett.116.061102,Abbott_2019,2020arXiv201014527A,PhysRevLett.119.161101,2017ApJ...848L..12A} is a vibrant field that constantly expands our understanding of gravitational phenomena and our universe. In the detection schemes currently employed by gravitational-wave detectors~\cite{Affeldt_2014,PhysRevLett.123.231108,PhysRevLett.123.231107}, such as KAGRA~\cite{10.1093/ptep/ptaa125}, GEO600~\cite{Dooley_2016}, Virgo~\cite{2015CQGra..32b4001A}, and LIGO~\cite{1992Sci...256..325A,2015CQGra..32g4001L}, the core algorithmic methods consists of excess energy-based burst searches~\citep{2000IJMPD...9..303A,2005CQGra..22S1159C,2001PhRvD..63d2003A,2002PhRvD..66f2002V,2004CQGra..21S1819K,2005CQGra..22S1311R,2008CQGra..25k4029K} and matched filtering searches
~\cite{1989thyg.book.....H,PhysRevD.60.022002,1993PhRvL..70.2984C,1994PhRvD..49.2658C,1998PhRvD..57.4535F,1998PhRvD..57.4566F,2004PhRvD..69l2001A,PyCBCSoft,2012PhRvD..85l2006A,2019PASP..131b4503B,2005PhRvD..71f2001A,2014PhRvD..90h2004D,2016CQGra..33u5004U,2018PhRvD..98b4050N,2017PhRvD..95d2001M,2019arXiv190108580S,2020PhRvD.101b2003H,76e97cf9801544919973534ed7028b6a,2016CQGra..33q5012A, nitz2017detecting}. Matched filtering is a classic signal processing technique, which computes the correlation of the time-delayed input signal with a bank of templates. While the matched filtering method has outstanding performance and statistical rigor, it potentially has room for improvement in terms of computational complexity \cite{yan2021generalized}. In fact, millions of templates are being used in the LIGO matched filtering pipeline, with the number still expanding with the scope of the search~\citep{2020arXiv201014527A}.

As we deepen our search for gravitational wave signals, the issue of computational efficiency (namely, the number of basic operations required by a computer) is becoming increasingly prominent. Detection methods that excel in both statistical performance and computational efficiency can significantly boost our capacities for exploring wider and higher-dimensional parameter spaces, and even other families of eccentric waveforms \cite{gayathri2022eccentricity}. This in turn will help with uncovering more astrophysical events, potentially unveiling novel astrophysical phenomena, as well as reducing the carbon footprint associated with searching for these events. 

In the literature, a promising approach to reducing the complexity of matched filtering searches has been to apply a two-step hierarchical search, which seeks to rapidly reject most negative samples \cite{mohanty1996hierarchical}.
Later, \cite{sengupta2003faster} expands the hierarchy to involve temporal multi-scale approach. Some other meritorious extensions include using geometric template placing \cite{owen1999matched} and hierarchy based on chirp times \cite{sengupta2002extended}. The work of \cite{gadre2019hierarchical} applies two-step detection within the PyCBC framework, and compares the performance on simulated data. A recent work of \cite{dhurkunde2021hierarchical} further combines the two-step method with dimensionality reduction in the template space using principle component analysis (PCA). All the above examples demonstrate improvements relative to basic matched filtering in various settings for GW detection.

Similar ideas have also been widely explored and applied in machine learning contexts. For example, \cite{grauman2005pyramid, chen2012hierarchical} consider hierarchical matching of image features in both spatial domain and feature domain for image classification. \cite{lazebnik2006beyond} considers image classification using hierarchical matching in the spatial domain. In natural language processing, hierarchical model have also been used in sentiment classification \cite{shen2018sentiment}. More specific applications of this idea include medical imaging \cite{vaillant1999hierarchical}, human detection and segmentation \cite{lin2007hierarchical}, and crime classification \cite{wang2019hierarchical}.

In the meantime, with the growing literature of applying deep learning and neural networks on GW detection, it is tempting to leverage deep learning's power to reduce complexity. Indeed, various neural network architectures have been shown to perform tasks such as GW detection, parameter estimation, noise transients identification and data denoising 
~\cite{gebhard2017convwave, george2018deep, gabbard2018matching, george2018deepneural, fan2019applying, morawski2020convolutional, dreissigacker2019deep, krastev2020real, lin2020binary, lin2020detection, bresten2019detection, astone2018new, yamamoto2020use, dreissigacker2020deep, corizzo2020scalable, miller2019effective, bayley2020robust, krastev2020detection, luo2020extraction, santos2020gravitational, chan2020detection, xia2021improved,biswas2013application, george2017deep, mukund2017transient, razzano2018image, fan2019applying, coughlin2019classifying, colgan2020efficient,nakano2019comparison, green2020gravitational, marulanda2020deep, caramete2020characterization, delaunoy2020lightning,shen2019denoising, wei2020gravitational,gebhard2019convolutional},
at performance levels comparable to that of matched filtering. Furthermore, it has been shown that matched filtering is generally suboptimal for parametric signal detection \cite{dent2014optimizing, yan2021generalized}, and the performance can bee improved by optimizing the templates using deep learning techniques \cite{yan2021generalized}. This can be achieved by setting up a neural network that is formally equivalent to matched filtering, and then training on data. Inspired by the flexibility of deep learning models, it is conceptually appealing to explicitly incorporate computational efficiency into the neural network objectives, aim to achieve  ``the best of both worlds.''

In this work, we propose a novel neural network architecture, named Hierarchical Detection Network (HDN), which takes the form of a multi-layer matched filtering with trainable parameters. In order to achieve the dual goal of accuracy and efficiency, we constructed a \textit{novel loss function} that explicitly incorporates computational complexity. We demonstrate the efficiency gains on data with open LIGO noise data and synthetic GW signal injections. As a quick glance at the performance gains, when tested on synthetically injected data at $\mathrm{SNR}=9$, compared with matched filtering, two-layer HDN can achieve false positive and false negative rates $0.2\%$ with $79\%$ lower complexity, and reduces error rates by $88\%$ when at equal complexity equivalent to 100 templates, for instance. Experimental details are described in Section \ref{sec:experiments}.

Yet, the two-layer networks do not reveal the full power of the proposed model. We further show that by training a three-layer model with careful initialization, it is capable of achieving even better accuracy at lower complexity. We also provide some intuitive insights into the mechanism behind multi-layer hierarchical models and their construction.

The rest of the paper is organized as follows. Section \ref{sec:preliminaries} reviews the problem of parametric detection and some relevant models. Section \ref{sec:main-hdn} introduces Hierarchical Detection Networks, including the setup, complexity and training process. Section \ref{sec:reduction} further discusses the complexity reduction from HDN. Section \ref{sec:experiments} presents experimental results of applying HDN on real LIGO data and synthetic injections. We discuss some further implications and future steps of this work in Section \ref{sec:discussion}.

\section{Preliminaries}
\label{sec:preliminaries}

In this section, we describe the problem setup for a single gravitational-wave detector, and review some related detection algorithms, including matched filtering and a closely related model MNet \cite{yan2021generalized}. 

\subsection{Parametric Detection and Matched Filtering}

Consider the problem of detecting gravitational waves in a single gravitational-wave detector data stream. Formally, assume we observe detector data $\mb x\in \R^n$, and need to decide whether or not $\mb x$ contains any gravitational-wave signal in addition to noise. Throughout this paper, we model the inputs as having fixed dimension, where time-domain convolution can be applied for generic time-series inputs. The signals of interest are modeled as belonging to a parametric set
\begin{equation}
    S_\Gamma = \{ \mb s_{\mb \gamma} \mid \mb \gamma \in \Gamma \}
\end{equation}
where the parameters $\mb \gamma$ can represent properties of the objects that generate the gravitational wave, such as masses, orbits and spins. We model noise as a random vector $\mb z\in\R^n$ with distribution $\rho_0$ and independent of the signal. The detection problem can thus be written as the following hypothesis test:
\begin{eqnarray}
    &H_0:& \mb x = \mb z, \\
    \text{or} \quad &H_1:& \mb x = \mb s_{\mb \gamma} + \mb z \ \text{ for some } \ \mb  \gamma \in \Gamma,
\end{eqnarray}
For this testing problem, we need to design decision rules $\delta: \mathbb{R}^n \to \{ 0, 1 \}$ that ideally achieve both good statistical performance and high computational efficiency. 

One natural approach to this problem is matched filtering \cite{turin1960introduction, woodward2014probability, schutz1999gravitational}, a classical method for signal detection. A matched filter is the optimal linear filter for maximizing the signal-to-noise ratio (SNR) in stationary Gaussian white noise \cite{turin1960introduction}. When detecting a single target signal $\mb s$, matched filtering takes the following form of an inner product test
\begin{equation}
    \delta(\mb x)=1 \text{ iff } \left<\mb s, \mb x\right> \ge \tau,
\end{equation}
where $\tau$ is a cutoff threshold.
Its natural extension to detecting a parametric family of signals is to take the maximum over template samples $\mb s_{\mb \gamma_1},\dots,\mb s_{\mb \gamma_k}$ in the signal space, namely
\begin{equation}
    \delta(\mb x)=1 \text{ iff } \max_{i=1,\dots,k} \left<\mb s_{\mb \gamma_i}, \mb x\right> \ge \tau.
\end{equation}
As noted in the literature \cite{yan2021generalized}, matched filtering is suboptimal for the generic parametric detection problem in terms of statistical performance. To address such suboptimality, one alternative approach that has been described is the MNet architectures.

\subsection{MNet Architectures}

The MNet architectures, including MNet-Deep and MNet-Shallow, are two machine learning model for parametric detection based on neural networks \cite{yan2021generalized}. Formally, an MNet is a neural network initialized to exactly replicate matched filtering, and then trained on data for improved performance. Such improvement is mainly due to its structural capability of handling non-convex decision boundaries and non-Gaussian noise distributions. The shallow and deep versions differ in how the replication of matched filtering is constructed. MNet-Shallow is structured identically to matched filtering, allowing for a more direct comparison, whereas MNet-Deep uses a multi-layer pairwise-max structure for higher flexibility and statistical performance. 

In this work, with computational complexity in mind, we will focus on the MNet-Shallow model, which can be expressed as:
\begin{equation}
    f_{\text{MNet-Shallow}}(\mb x) = \max_i \left< \mb s_i, \mb x \right>.
\end{equation}
Here the weights $\mb s_i$ are initialized as templates and then trained over data, making it advantageous over classic matched filtering. If we compare its computational efficiency against matched filtering (measured in terms of the number of operations required to achieve a target error rate), the strict performance improvement with identical architecture suggests that one can expect a strict efficiency improvement as well.

However, the structural similarity between MNet-Shallow and matched filtering implies that such efficiency gains may typically be very limited. In order to achieve efficiency gains on higher orders of magnitude, we may need to reconsider the parametric detection problem, and innovate on the basic matched filtering rule. As we present in the next section, one solution is to arrange the templates in a multi-layer hierarchy, so that significant proportions of negative-labeled data are subject to early rejections.

\section{Hierarchical Detection Networks}
\label{sec:main-hdn}

In this section, we present the Hierarchical Detection Network (HDN), which improves over matched filtering and MNet-Shallow to simultaneously maximize statistical performance and computational efficiency. 

The main idea behind HDN is intuitive. If an input segment clearly contains no gravitational wave signals, we may not need to subject it to millions of templates to tell that. A small number of ``gatekeeper'' templates may be sufficient for confidently rejecting these ``obviously wrong'' instances. Once these inputs have been ruled out, we can apply a more refined test using possibly more templates, and reject a larger portion of the input space. This procedure can be repeated, until in the very last step, we employ our full template bank for a full diagnosis on the remaining instances which all previous tests failed to reject. Since the overwhelming majority part of the gravitational wave strain data contains noise only, most instances will likely be addressed by the initial simple layers of the model, saving the need for the full template bank. In addition, different layers of the HDN may be designed to specialize in different parts of the input space, such that the available parameter space of the potentially allowed waveforms are successively restricted as the hierarchical process progresses from later to layer, allowing for further efficiency gains.

\subsection{Architecture of HDN}

\begin{figure}[ht]
    \centering
    \includegraphics[width=.7\linewidth, trim={0 0.2cm 0 0.1cm}, clip=true]{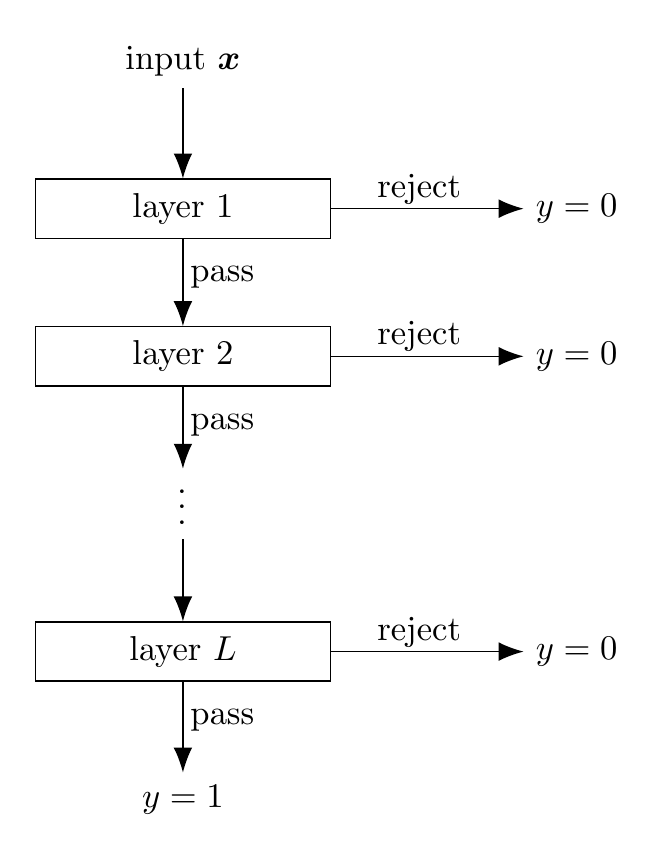}
    \caption{Illustration of a hierarchical detection network.}
    \label{fig:hierarchical}
\end{figure}

We first formally define a hierarchical detection network (HDN). Generally speaking, a HDN is a hierarchical template matching model trained as a neural network, as illustrated in Fig \ref{fig:hierarchical}. Let $L$ be the number of layers in the hierarchical structure, and let $\{\mb s_i\}_{i=1}^n$ be the entire set of $n$ templates used by the model. For each layer $\ell=1,\dots,L$, only the first $n_\ell$ of these templates are used in that layer, where $0 < n_1 < \dots < n_L = n$. Let the threshold associated with template $i$ at layer $\ell$ be $t_{i,\ell}$, $i=1,\dots,n_\ell$. Here we let layer $l$ reuse all templates from the previous layer(s), but assign independent threshold values to the reused templates at different layers, in order to reduce computation complexity.

Following conventions of the machine learning literature on binary classification, we call an input $\mb x$ \textit{positive} if it contains a GW signal, and \textit{negative} if it only contains noise. For a given input $\mb x$, the model processes it using the following procedure:

\begin{algorithm}[H]
\caption{The HDN algorithm}\label{alg:hdn}
\hspace*{\algorithmicindent} \textbf{Parameters:} $L, n, \{n_\ell\}, \{\mb s_i\}, \{t_{i,\ell}\}$ \\
\hspace*{\algorithmicindent} \textbf{Input:} $\mb x$
\begin{algorithmic}
\hrule
\State $\ell\gets 1$
\While {$\ell\le L$}
    \State $y_\ell \gets \max_{i\le n_\ell} \left< \mb x, \mb s_i \right> - t_{i,\ell}$
    \If {$y_\ell < 0$}
        \State \textbf{return} 0
    \EndIf
    \State $\ell\gets \ell+1$
\EndWhile
\State \textbf{return} 1
\end{algorithmic}
\end{algorithm}

More formally, for a given input $\mb x$, let
\begin{equation}
    y_\ell = \max_{i\le n_\ell} \left< \mb x, \mb s_i \right> - t_{i,\ell}
\end{equation}
be the matching output at layer $\ell$, $\ell=1,\dots,L$. Let
\begin{equation}
    I_\ell = \begin{cases}
    \mathbbm{1}[y_{\ell}>0], &\text{ if } I_{\ell-1}=1 \\
    0, &\text{ if } I_{\ell-1}=0
    \end{cases}
    \label{eqn:indicators}
\end{equation}
be the indicator of whether the input passes layer $\ell$ of the model on to the later layer(s), $\ell=1,\dots,L-1$, and define $I_L\equiv0$. With these notations, the overall output of the model can be written as
\begin{equation}
    y(\mb x) = \sum_{\ell=1}^{L} y_\ell (1-I_\ell) \prod_{k=1}^{\ell-1} I_k.
    \label{eqn:y-precise}
\end{equation}

Note that both matched filtering and MNet-Shallow can be unified under the framework of HDN, viewed as a model with a single layer. In the meantime, some of the existing two-step MF methods \cite{mohanty1996hierarchical, owen1999matched, gadre2019hierarchical} can also be interpreted under this framework. Furthermore, the HDN architecture is not restricted to the typical two-layer hierarchy of coarse and fine searches, but can utilize multiple layers which specialize in different parts of the signal space. The use of multiple layers and their setup is further discussed in Section \ref{sec:reduction}.

\subsection{Measure of Computational Complexity}

With matched filtering and HDN unified under the same framework, we can provide a formal definition of computational complexity to facilitate our discussion. We are often most concerned about the execution efficiency of the model in deployment rather than in training, since it determines the real-time processing abilities. In the meantime, any computational cost of setting up the parameters of the model, including template selection for matched filtering and training for neural networks, is a one-time cost and can be conducted offline. Therefore it is natural to define complexity based on test time.

Also, since for the vast majority of time the input strain does not contain gravitational wave events, we can capture the computational complexity solely by its performance on negative data. This leads to the following definition of complexity:
\begin{definition}[Complexity]
The (computational) complexity of a HDN model is defined as the expected number of template matching (inner product) operations conducted to evaluate a negative input.
\end{definition}

Formally, we can write the complexity as 
\begin{equation}
    Z = \E_{\mb x\sim F_-}[z(\mb x)],
\end{equation}
where
\begin{equation}
    z(\mb x) = \sum_{\ell=1}^{L} n_\ell (1-I_\ell) \prod_{k=1}^{\ell-1} I_k
    \label{eqn:z-precise}
\end{equation}
is the number of matching operations required for evaluating an input $\mb x$.

To illustrate this measure of complexity, note that for matched filtering and MNet-Shallow models, the complexity simply equals the number of templates used in the model. For a two-layer HDN, assuming only a proportion $p$ of negative data enters the second layer, the complexity for the model will be $n_1 + p\cdot n_2$. Intuitively, if the initial layer contains fewer templates while being able to reject a significant portion of negative inputs, these inputs will not need to undergo the entire model, hence reducing the complexity of the model. This straightforward idea forms the basis of HDN, upon which we further leverage the power of data through training for an additional boost in performance.

\subsection{Training of HDN}

So far, we have described the behavior of HDN at test/deployment time, and now we turn our attention to the training process. Conceptually, we want to set up a loss function as an appropriate combination of classification error and model complexity, so that minimizing the loss would achieve simultaneously accuracy and efficiency. However, a loss function directly based on the above expressions \eqref{eqn:y-precise} and \eqref{eqn:z-precise} is undesirable because of non-differentiability. Instead, we use soft surrogates for the indicators $I_\ell$. Define
\begin{equation}
    \hat{I}_\ell = \phi(y_\ell)
\end{equation}
for $\ell=1,\dots,L-1$ where $\phi(x):= \frac{1}{1+e^{-x}}$ is the sigmoid function, serving as a soft surrogate of the step function. Also let $\hat{I}_L \equiv 0$. Note that during training we can simply compute $y_\ell$ for all layers regardless of whether previous layers were passed, since this will only be a one-time offline cost. Define the soft surrogates for $y(\mb x)$ and $z(\mb x)$ accordingly:
\begin{align}
    \hat{y}(\mb x) &= \sum_{\ell=1}^{L} y_\ell (1-\hat{I}_\ell) \prod_{k=1}^{\ell-1} \hat{I}_k, \label{eqn:y-approx} \\
    \hat{z}(\mb x) &= \sum_{\ell=1}^{L} n_\ell (1-\hat{I}_\ell) \prod_{k=1}^{\ell-1} \hat{I}_k. \label{eqn:z-approx}
\end{align}

Assume the training dataset is $\{(\mb x_i, y^\star_i)\}_{i=1}^N$, with $N_+$ positive entries and $N_-$ negative entries. The loss function can be formulated as
\begin{align}
    \mathcal{L} = \frac{1}{N} \sum_i \ell_i^{\mr{accu}} + \lambda\cdot \frac{1}{N_-} \sum_{i:y^\star_i=0} \ell_i^{\mr{cplx}},
    \label{eqn:loss-fn}
\end{align}
where 
\begin{equation}
    \ell_i^{\mr{accu}} = y_i^\star \log p_i + (1-y_i^\star) \log (1-p_i)
\end{equation}
with $p_i = \frac{1}{1+e^{-\hat{y}_i}}$ is equivalent to the cross-entropy loss for binary classification, and
\begin{equation}
    \ell_i^{\mr{cplx}} = \hat{z}_i
\end{equation}
is the soft approximate for the complexity of evaluate the negative inputs.
With the loss function \eqref{eqn:loss-fn} defined above, we can then train the model parameters $\{\mb s_i\}$ and $\{t_{i,\ell}\}$ using first order optimization methods. 

Experimental results of the HDN architecture will be shown in Section \ref{sec:experiments}.

\section{Complexity Reduction from Multiple Layers}
\label{sec:reduction}

Here we provide some theoretical insights into why the hierarchical model achieves reduced complexity at similar target performance levels, particularly with more layers.

Consider as an example a two-layer hierarchical model with $n_1, n_2$ templates respectively on the layers. Let $\alpha_\ell$ and $\beta_\ell$ denote respectively the FPR and FNR of layer $\ell$ conditioned on data that reaches the corresponding layer. The overall FPR, FNR and the complexity $z$ can then be represented as:
\begin{align}
    \alpha_{\text{all}} &= \alpha_1 \alpha_2 \\
    \beta_{\text{all}} &= \beta_1 + (1-\beta_1) \beta_2 \\
    z_{\text{all}} &= n_1 + \alpha_1 (n_2-n_1)
\end{align}

\begin{figure}[ht]
    \centering
    \includegraphics[width=3in, trim={0.4cm 0.6cm 0 1.2cm}, clip=true]{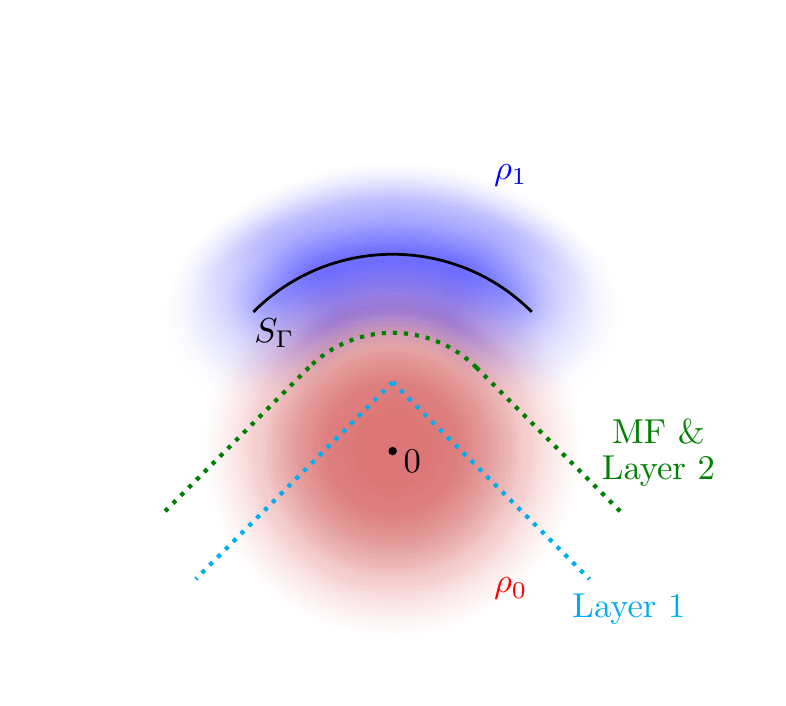}
    \caption{An example of the complexity advantage of hierarchical detection models.}
    \label{fig:cplx-example}
\end{figure}

To understand why an improvement in complexity can be expected, we consider the following example setup of parametric detection as shown in FIG. \ref{fig:cplx-example}. The probability density of the two labeled classes are $\rho_0$ and $\rho_1$ respectively. 
Imagine a baseline MF model with decision boundary as shown by the green curve, at the cost of $n$ templates, where $n$ has to be relatively large to approximate the smoothly curved boundary. Then we can construct the following hierarchical model to achieve a significantly lower complexity with identical statistical accuracy. To do this, we construct a simple two-layer hierarchical model, with the first layer decision boundary as shown by the dotted blue line, and the second layer decision boundary coinciding with that of the MF model. Notice that the first layer features a very low complexity $n_1$ (with $n_1=2$ in this example), and in the meantime has a fairly high true negative rate $1-\alpha$. Since the second layer reproduces the MF decision boundary, the overall decision rule of the hierarchical model is identical to that of the MF model, and hence they share exactly the same ROC (receiver operating characteristic) curves. However, the complexity of the HDN model is $n_1 + \alpha_1(n_2-n_1)$, which is significantly smaller than $n_2$ provided $n_1$ is small compared with $n_2$ and $\alpha_1$ is not too close to 1.

\begin{figure}[ht]
    \centering
    \includegraphics[width=3in, trim={0 0.5cm 0 1.2cm}, clip=true]{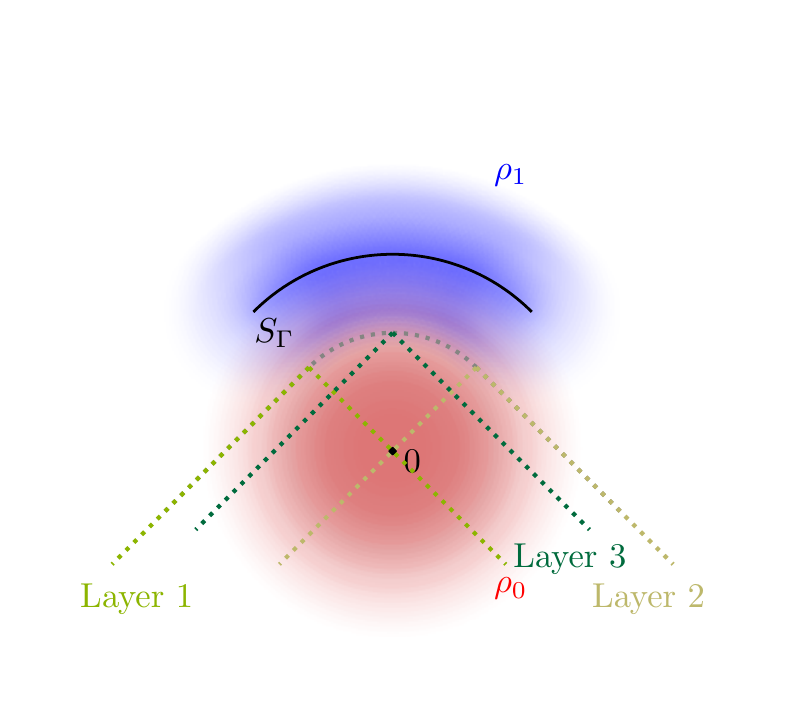}
    \caption{A hierarchical model with more simple layers that lie inside the overall negative decision region.}
    \label{fig:cplx-non-nested}
\end{figure}

This example provides inspirations for a general recipe for designing hierarchical models with reduced complexity. For any decision rule given by a MF model, we can construct a sequence of preceding layers whose negative decision regions all lie inside the negative decision region of the MF, and finally let the very last layer be equivalent to the original MF. The resulting hierarchical model will again have exactly the same overall decision rule and hence ROC curve, but with significantly reduced complexity. An illustrating example is shown in FIG. \ref{fig:cplx-non-nested}.

More generally, such constructions of hierarchical models can serve as good initializations for a HDN. One practical initialization scheme for an $L$-layer HDNs is the following: first train a separate $L-1$-layer model that only consists of the latter $L-1$ layers of the desired model. Then we initialize the latter $L-1$ layers of the original model with the trained network, and initialize the first layer with small $t_{i,1}$ values such that almost all inputs pass. This gives an initialization of the $L$-layer model which at initialization essentially replicates the $L-1$-layer model. From there, we train the initialized $L$-layer model on data, which will leverage the higher architectural capacity for further improved performance and complexity. An experiment that illustrates this approach is shown in Section \ref{sec:experiments}.

\section{Simulation and Experiments}
\label{sec:experiments}

\subsection{Data Generation}

In the experiments, we use open L1 strain data from LIGO Livingston's O2 run between August 1 and August 25, 2017 with ANALYSIS{\textunderscore}READY flag \cite{GWOSCanalysis}. The total duration of the frame files is $389.12$ hours. We downsample the strain data from the original $4096$Hz to $2048$Hz for processing efficiency. The downsampled data is then divided into segments of 2 seconds, with each segment overlapping with $50\%$ of its preceding segment. 

To evaluate the accuracy of detection models, we need both positive and negative datasets. For the negative datasets, the strain data itself is used. For the positive datasets, due to the very limited number of confirmed detections of GW events, we generate positive data by injecting synthetic waveforms into the noise strains, at a preset SNR value.

The entire L1 strain dataset is first divided into two sets to be used in training and test respectively, such that any segment in the training set does not overlap with any segment in the test set. For training and test respectively, a positive and a negative dataset are generated. For the positive datasets, synthetic waveforms are generated with masses $m_1,m_2$ uniformly drawn from $[20,50]$ and 3-dimensional spins drawn from an isotropic distribution and with spin dimensionless magnitudes drawn from a uniform distribution within [0,1]. The injected waveforms are aligned such that the peak is located at 0.95 second, and the injection amplitude is chosen such that the signal-to-noise ratio (SNR) after preprocessing is constant at 9. The preprocessing is applied to all data (after injection if applicable) by using an FIR bandpass filter with cutoff frequency $30$Hz and $400$Hz, whitening with power spectral density estimated from the L1 strain data, and truncating to only keep the center 1 second.

\subsection{Two-Layer Networks}

In this experiment, we limit our HDN models to two layers and $n_2 = 10n_1$. At initialization, the templates $\mb s_i$ are chosen as random gravitational waveforms from the same parameter space, and the thresholds $t_{i,\ell}$ are set to the same within each of the two layers. The parameter $\lambda$ is the loss function is fixed at $\lambda=10^{-4}$. For the optimization procedure of the network, we use the Adam optimizer and a constant learning rate of $10^{-4}$.

\begin{figure}[ht]
    \centering
    \includegraphics[width=.9\linewidth, trim={1cm 0.5cm 0.5cm 0.5cm}]{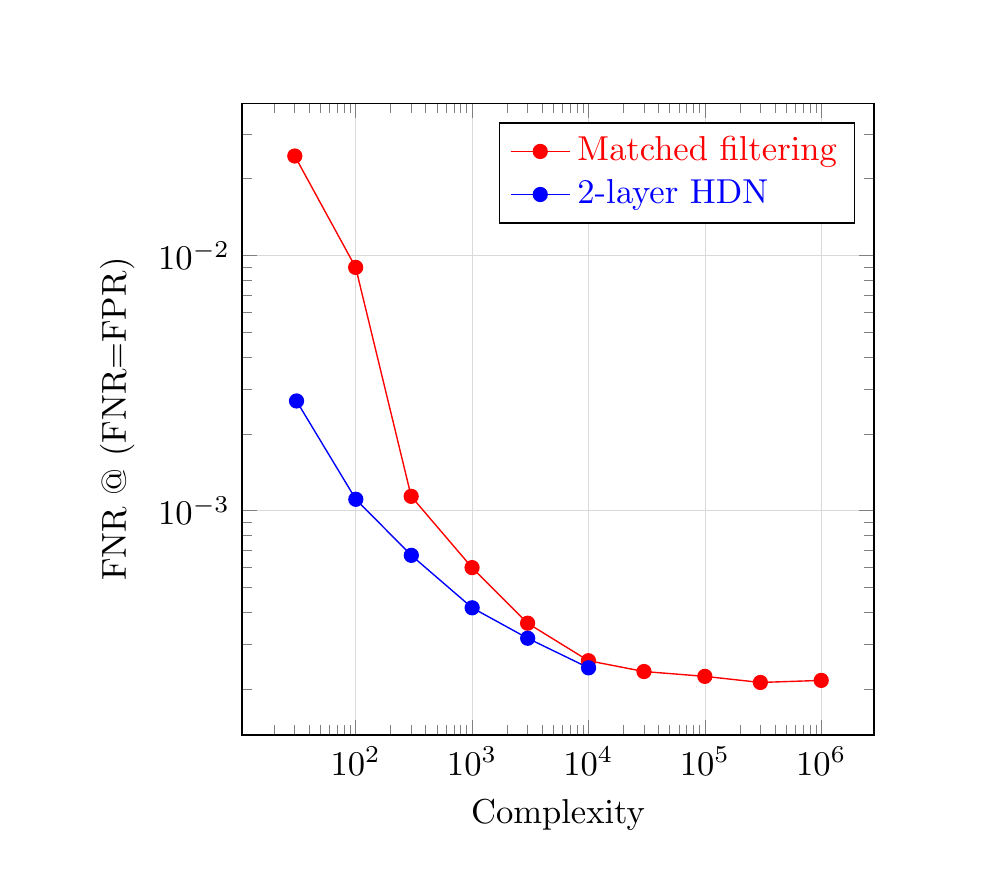}
    \caption{Complexity-performance trade-off of matched filtering and the hierarchical neural network.}
    \label{fig:complexity}
\end{figure}
\begin{figure}[ht]
    \centering
    \includegraphics[width=.9\linewidth, trim={0.5cm 0.5cm 0.5cm 4cm}]{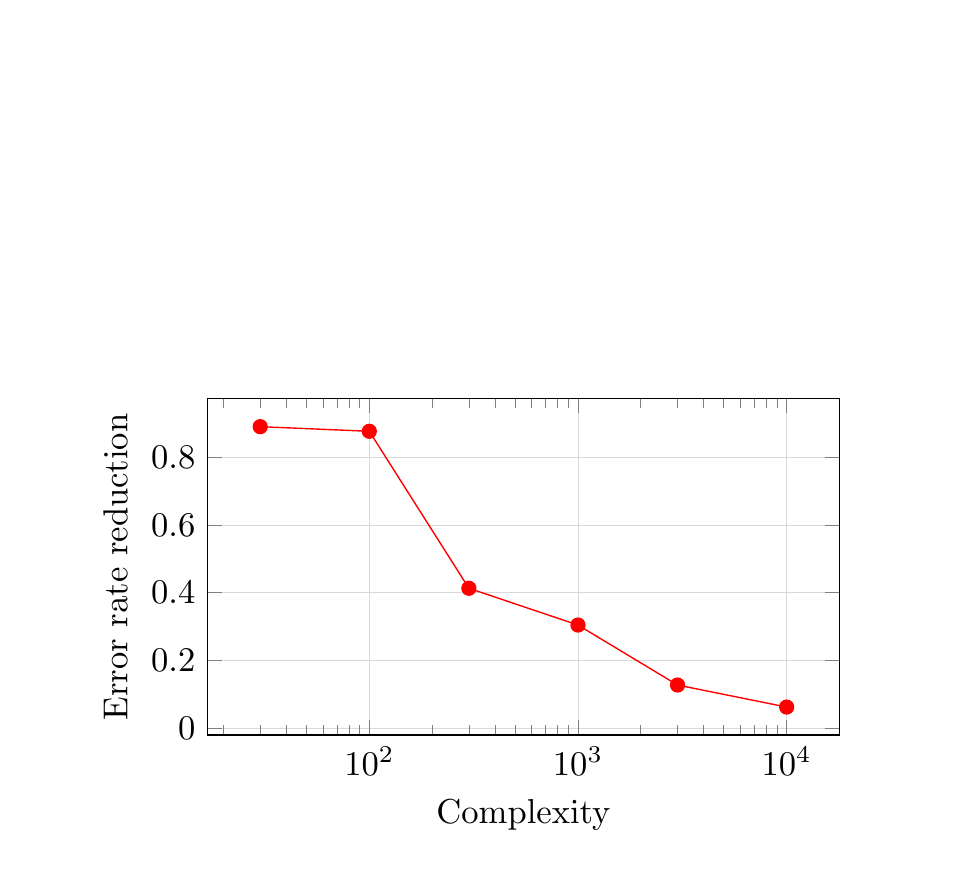}
    \caption{Proportion of error rate reduced by using HDN over matched filtering.}
    \label{fig:complexity-ratio}
\end{figure}

FIG. \ref{fig:complexity} shows the comparison of the complexity-performance trade-offs of MF and HDN models, where the HDN models are two-layer architectures structured as described above. The horizontal axis plots the logarithm of the complexity measure defined in this paper, and the vertical axis plots the logarithm of error rates at the point on the ROC curve where $\fpr=\fnr$. This choice of measure eliminates the arbitrariness of choosing FNR at a fixed FPR level. For each architecture, 10 independent runs are conducted, and the one with lowest accuracy measure is shown. The red curve for HDN is cut off early due to memory limitations of training the model. We see that HDN consistently achieves a lower complexity than MF at equal accuracy.

\subsection{Three-Layer Networks}

We further demonstrate the power of the proposed model with a deeper three-layer network. Conceptually, since adding more layers strictly improves model expressability, it should never hurt performance provided that the parameters are initialized or trained appropriately. 

\begin{figure}[ht]
    \centering
    \includegraphics[width=.98\linewidth, trim={0 0 0 0}]{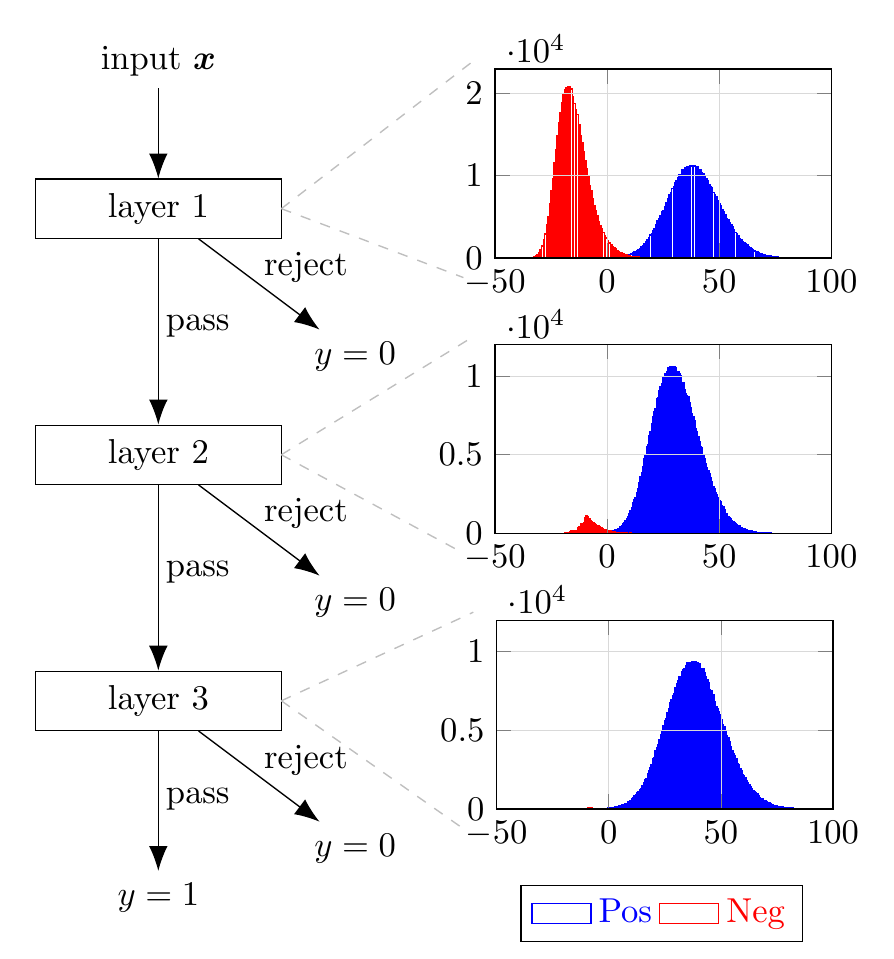}
    \caption{Illustration of the 3-layer architecture, and the output densities on the test data from each layer. Only data entries that reach a given layer is shown. We see that each layer successfully rejects the vast majority of incoming negative data, and barely any negative data reaches the last layer.}
    \label{fig:3layer-density}
\end{figure}

In this experiment, we construct a 3-layer HDN with layer sizes $(n_1,n_2,n_3) = (30, 100, 1000)$ in the following way. First, a shallower 2-layer model with layer sizes $(100, 1000)$ is trained, and we use these trained parameters to initialize the latter two layers of the 3-layer model. We then initialize the first layer of the 3-layer model, setting the per-template thresholds $t_{i,1}$ as the same value for all $i$, such that all training data passes this first layer at initialization. This scheme ensures that at initialization, the 3-layer model essentially replicates the performance of the trained 2-layer model, giving it a head start before entering the training phase. The training is done in the same way as described before. FIG. \ref{fig:3layer-density} illustrates the architecture of this 3-layer model, along with the output densities of test data that reaches that layer, separated by the true class labels. Namely, the densities of layer 1 involves all input data, and the densities of layers 2 and 3 involve only the data that pass the previous layers. We see that most of the negative data are successfully intercepted by the initial layers, with very few of them reaches the final layer, which corroborates our intuition.

Here when evaluating the ROC curve, we adopt a slightly different approach that is more consistent with deep hierarchical models. Notice from equation \eqref{eqn:indicators} that the model uses a built-in threshold 0 to control the passing of each layer. When generating the ROC curve using a varying threshold, such a threshold should be applied at all layers instead of only the last layer. Therefore, at test time only, we replace the threshold 0 in equation \eqref{eqn:indicators} with a variable threshold $t\in\R$ which is constant for all layers, and compute the test outputs using \eqref{eqn:y-precise} as before for each $t$ value. Varying this threshold $t$ produces the ROC curve. Also note that $t$ determines which test entries would pass the layers, hence it also affects the model complexity evaluated on the negative test dataset. In actual deployment, the threshold $t$ should be fixed at some level that gives the desired trade-off between FPR and FNR, so this is only for demonstration purpose.

\begin{figure}[ht]
    \centering
    \includegraphics[width=.9\linewidth, trim={2cm 0 0 0}]{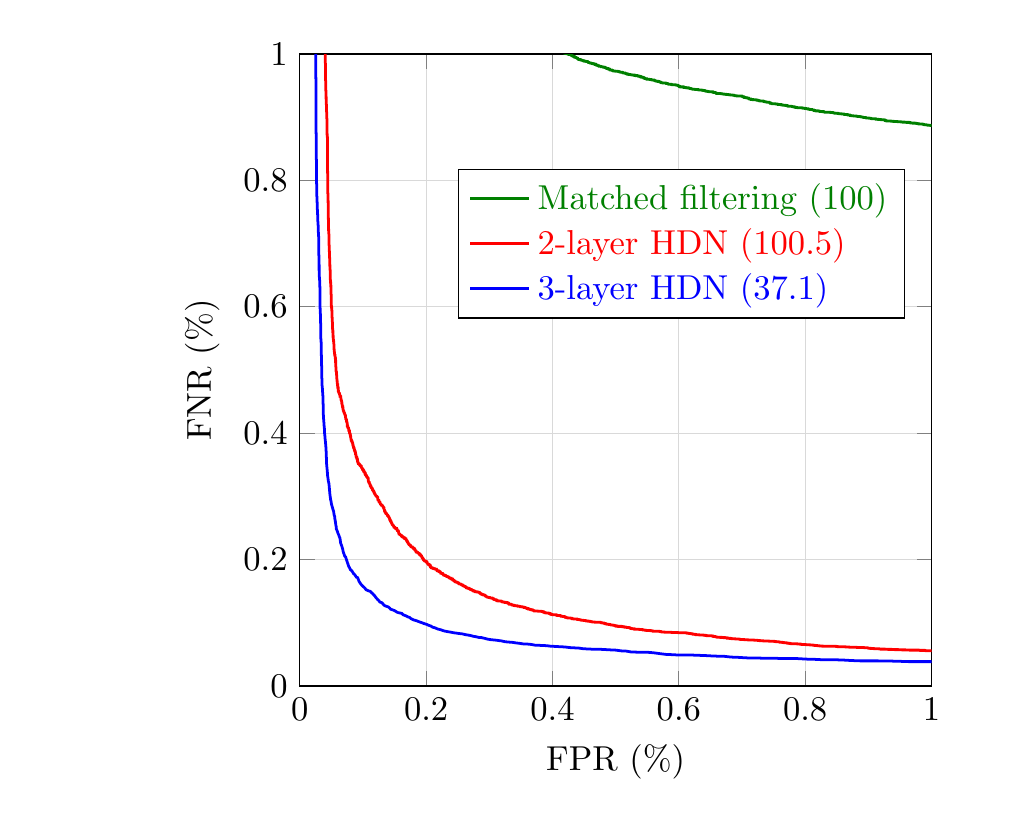}
    \caption{Comparison of ROC curves between three models. The numbers in parentheses show the complexity of the model.}
    \label{fig:3layer-roc}
\end{figure}

FIG. \ref{fig:3layer-roc} shows the comparison of ROC curves between a matched filtering model with 100 templates, a 2-layer HDN model with complexity $100.5$ (used for initializing the 3-layer model), and a 3-layer HDN model with complexity $37.1$ at the point of equal FPR and FNR. While the complexity of the 3-layer model depends on the specific point chosen on the ROC curve, it does not exceed 65 for the entire segment of ROC curve shown in the figure, and is thus always lower than the 2-layer model. We see that the deeper 3-layer model excels at both accuracy and efficiency compared with the 2-layer model, and significantly more so if compared with the matched filtering model. This further showcases the power of depth in hierarchical models, and corroborates our discussion in Section \ref{sec:reduction}.

\section{Discussion}
\label{sec:discussion}

In this paper, we showed that by leveraging ideas from classical matched filtering and modern machine learning, we are able to design systems for GW detection that simultaneously optimize statistical accuracy and efficiency. This general conceptual idea of trainable hierarchical matched filtering can be applied upon a wide range of existing proposals for efficient detection pipelines.

While the proposed HDN model conducts hierarchical rejection on the data, an alternative can be proposed to conduct hierarchical acceptance, namely to progressively label parts of the data as positive rather than negative. This has the advantage of aligning better with the matched filtering routine, since it suffices to use one matching template to confirm a signal. In the specific problem of GW signal detection, due to the class imbalance from the scarcity of actual GW events, the majority of computational complexity hinges on the classification of negative data, and therefore a hierarchical rejection model will have much more significant efficiency gains. In more general signal detection problems, hierarchical acceptance constructions can also be deployed in similar fashions as HDN.

The proposed HDN can potentially have wider applications within the field of gravitational wave science. For example, in the task of glitch detection and identification 
~\cite{biswas2013application,cavaglia2018finding,2020PhRvD.101j2003C,2022arXiv220213486C,2022arXiv220305086C,Cuoco_2020},
one can combine existing constructions of machine learning based models with hierarchical models, to improve on both efficiency and accuracy.

One aspect to be further explored is how to select the number of layers in the hierarchy. While having more layers can boost model expressability and further leverage the efficiency gains, excessive hierarchy may offer diminishing returns, and also make training increasingly difficult. In the paper we demonstrated how a 3-layer model excels over a 2-layer one, and there may be a ``sweet spot'' number of layers for a given signal detection setups. Another promising direction would be to incorporate prior knowledge about the signal domain such as low-dimensionality and representative features into the detection model, which may be able to further outperform these current models agnostic of the signal space properties.

\begin{acknowledgments}
This material is based upon work supported by NSF's LIGO Laboratory which is a major facility fully funded by the National Science Foundation.

We acknowledge computing resources from Columbia University's Shared Research Computing Facility project, which is supported by NIH Research Facility Improvement Grant 1G20RR030893-01, and associated funds from the New York State Empire State Development, Division of Science Technology and Innovation (NYSTAR) Contract C090171, both awarded April 15, 2010.

The authors are grateful for the LIGO Scientific Collaboration for the careful review of the paper. This paper is assigned a LIGO DCC number of LIGO-LIGO-P2200203. 
The authors acknowledge the LIGO Laboratory and Scientific Collaboration for the detectors, data, and the game changing computing resources (National Science Foundation Grants PHY-0757058 and PHY-0823459). The authors would like to thank colleagues of the LIGO Scientific Collaboration and the Virgo Collaboration and Columbia University for their help and useful comments, in particular the CBC group, Andrew Williamson, Stefan Countryman, William Tse, Nicolas Beltran, Asif Mallik, Sireesh Gururaja, and Thomas Dent which we hereby gratefully acknowledge.
SM thanks David Spergel, Rainer Weiss, Rana Adhikari, and Kipp Canon for the motivating general discussions related to the role of machine learning and data analysis. 

This research has made use of data, software and/or web tools obtained from the Gravitational Wave Open Science Center~\cite{ABBOTT2021100658,GWOSC} (https://www.gw-openscience.org/ ), a service of LIGO Laboratory, the LIGO Scientific Collaboration and the Virgo Collaboration. This material is in part based upon work supported by NSF’s LIGO Laboratory which is a major facility fully funded by the National Science Foundation, specifically some of the testing used LIGO detector noise publicly released by the LIGO Scientific Collaboration. LIGO Laboratory and Advanced LIGO are funded by the United States National Science Foundation (NSF) as well as the Science and Technology Facilities Council (STFC) of the United Kingdom, the Max-Planck-Society (MPS), and the State of Niedersachsen/Germany for support of the construction of Advanced LIGO and construction and operation of the GEO600 detector. Additional support for Advanced LIGO was provided by the Australian Research Council. Virgo is funded, through the European Gravitational Observatory (EGO), by the French Centre National de Recherche Scientifique (CNRS), the Italian Istituto Nazionale di Fisica Nucleare (INFN) and the Dutch Nikhef, with contributions by institutions from Belgium, Germany, Greece, Hungary, Ireland, Japan, Monaco, Poland, Portugal, Spain.

The authors thank the University of Florida and Columbia University in the City of New York for their generous support.The authors are grateful for the generous support of the National Science Foundation under grant CCF-1740391. The authors thank Sharon Sputz of Columbia University for her effort in facilitating this collaboration. 

I.B. acknowledges the support of the National Science Foundation under grant PHY-1911796 and the Alfred P. Sloan Foundation.

\end{acknowledgments}

\bibliographystyle{unsrt}
\bibliography{ref,EquivalencePaperRefs}

\appendix

\end{document}